\documentclass[dvips]{article}
\usepackage{wasysym}
\usepackage{icrctc07}

\title{Observations of microquasars with the MAGIC telescope}
\shorttitle{Observations of microquasars with the MAGIC telescope}

\authors{Javier Rico$^{1,*}$ Michael Rissi$^{2}$, Pol Bordas$^{3}$,
Valenti Bosch-Ramon$^{3}$, Juan Cortina$^{1}$, Josep M. Paredes$^{3}$,
Marc Rib\'o$^{3}$, Diego F. Torres$^{4}$ and Roberta Zanin$^{1}$ for
the MAGIC Collaboration} 
\shortauthors{J. Rico et al.}
\afiliations{$^1$Institut de Fisica d'Altes Energies, Barcelona, Spain\\ 
$^2$Institute for Particle Physics, ETH Zurich, Switzerland\\
$^3$Departament d'Astronomia y Meteorologia, Universitat de
Barcelona, Barcelona, Spain\\
$^4$ICREA \& Institut de Ciencies de l'Espai (IEEC-CSIC), Barcelona, Spain
}
\email{$^*$Presenter (jrico@ifae.es)}

\abstract{We report on the results from the observations in very high energy
band (VHE, $E_\gamma \ge 100$ GeV) of the black hole X-ray binary
(BHXB) Cygnus X-1. The observations were performed with the MAGIC
telescope, for a total of 40 hours during 26 nights, spanning the
period between June and November 2006. We report on the results of the
searches for steady and variable $\gamma$-ray signals, including the
first experimental evidence for an intense flare, of duration between
1.5 and 24 hours.}

\begin{document}
\maketitle

Cygnus X-1 is the best established candidate for a stellar mass
black-hole (BH) and one of the brightest X-ray sources in the sky
\cite{Bowyer1965}. Located at a distance of $2.2\pm 0.2$~kpc, it is composed
of a $21\pm 8$~M$_\odot$ BH orbiting an O9.7~Iab companion of $40\pm
10$~M$_\odot$ \cite{Ziolkowski2005} in a circular orbit of 5.6 days
and inclination between 25$^\circ$ and 65$^\circ$ \cite{Gies1986}. The
X-ray source is likely powered by accretion and displays the canonical
high/soft and low/hard X-ray spectral states. The thermal soft
component is produced by the accretion disk close to the BH, whereas
hard X-rays are thought to be produced by inverse Compton scattering
of soft photons by thermal electrons in a corona or at the base of a
relativistic jet.
 
Observations in the soft $\gamma$-ray range with COMPTEL
\cite{Mcconnell2002} and {\it INTEGRAL} \cite{Cadolle2006}
strongly suggest the presence of a higher energy, non-thermal
component. Images with the VLBA have shown the presence of a highly
collimated relativistic jet
\cite{Stirling2001}. Cygnus X-1 could be a ``microblazar'',
where the jet axis is roughly aligned with the line of sight
\cite{Romero2002}. The interaction of the jet with
the interstellar medium appears to be responsible for a large-scale
($\sim 5$ pc diameter), ring-like, radio emitting structure
\cite{Gallo2005}.

\begin{table}[!th]
\footnotesize
\begin{center}
\begin{tabular}{rrrrr}
\hline
 MJD & $T$ & $N_\textrm{excess}$ & $S$ & Post \\ $[$days$]$& [min] &
 [evts] &[$\sigma$] & [$\sigma$] \\
\hline
 53942.051&     61.1&     3.6$\pm$  4.8&     0.8&     $<0.1$\\
 53964.887&    105.6&     4.8$\pm$  6.9&     0.7&     $<0.1$\\
 53965.895&    195.3&   $-$13.2$\pm$ 10.1&    $-$1.3&     $<0.1$\\
 53966.934&    124.8&     9.4$\pm$  9.5&     1.0&     $<0.1$\\
 53967.992&     48.5&    $-$9.0$\pm$  4.7&    $-$1.7&     $<0.1$\\
 53968.883&    237.5&    $-$4.4$\pm$ 11.6&    $-$0.4&     $<0.1$\\
 53994.953&     53.6&    $-$4.0$\pm$  4.9&    $-$0.8&     $<0.1$\\
 53995.961&     58.1&    $-$2.8$\pm$  4.6&    $-$0.6&     $<0.1$\\
 53996.855&    176.2&     1.6$\pm$  9.1&     0.2&     $<0.1$\\
 53997.883&    132.7&     5.2$\pm$  7.6&     0.7&     $<0.1$\\
 54000.852&    165.2&    11.4$\pm$  9.7&     1.2&     $<0.1$\\
 54002.875&    154.4&    36.8$\pm$ 10.4&     4.0&     3.2\\
 54003.859&    166.9&    $-$7.0$\pm$  9.1&    $-$0.8&     $<0.1$\\
 54004.891&    123.3&    $-$6.0$\pm$  7.9&    $-$0.7&     $<0.1$\\
 54005.914&     87.9&    $-$2.2$\pm$  6.3&    $-$0.3&     $<0.1$\\
 54006.938&     28.0&     5.4$\pm$  4.1&     1.4&     $<0.1$\\
 54020.891&     65.5&    $-$8.6$\pm$  5.9&    $-$1.4&     $<0.1$\\
 54021.887&     68.6&    $-$6.2$\pm$  5.7&    $-$1.0&     $<0.1$\\
 54022.887&     58.1&     1.6$\pm$  5.9&     0.3&     $<0.1$\\
 54028.863&     68.6&     3.4$\pm$  5.9&     0.6&     $<0.1$\\
 54029.895&     33.5&     3.4$\pm$  5.1&     0.7&     $<0.1$\\
 54030.863&     19.6&    $-$1.8$\pm$  3.0&    $-$0.6&     $<0.1$\\
 54048.824&     47.2&     1.6$\pm$  5.7&     0.3&     $<0.1$\\
 54049.824&     47.9&    $-$6.0$\pm$  5.4&    $-$1.1&     $<0.1$\\
 54056.820&     27.1&    $-$5.2$\pm$  3.8&    $-$1.3&     $<0.1$\\
 54057.820&     21.5&     1.2$\pm$  2.6&     0.5&     $<0.1$\\
\hline
\end{tabular}
\caption{From left to right: Modified Julian Date of the
beginning of the observation, total effective observation time (EOT),
number of excess events, statistical significance of the excess,
equivalent (\emph{post-trial}) significance for 26 independent
samples. A cut SIZE$>$200 photo-electrons ($E_\gamma > 150$ GeV) has
been applied.\label{table:log}}
\end{center}
\end{table}

Three other binary systems have been detected so far in the VHE
domain: PSR~B1259$-$63 \cite{Aharonian2005a}, LS~I+61~303
\cite{Albert2006a} and LS~5039 \cite{Aharonian2005b}.
In PSR~B1259$-$63 the TeV emission is thought to be produced by the
interaction of the relativistic wind from a young pulsar with the
outflow of the companion star. Recent results suggest that
LS~I~+61~303 also contains a non-accreting neutron star
\cite{Dhawan2006}, while the situation is not yet clear for
of LS~5039 \cite{Khangulyan2007,Sierpowska2007}.  To date, however,
there is no experimental evidence of VHE emission from any galactic
BHXB system.

\begin{figure}[!t]
\begin{center}
\includegraphics [width=0.4\textwidth]{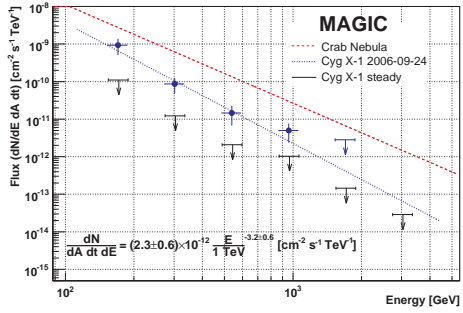}
\end{center}
\vspace{-0.5cm} 
\caption{Differential energy spectrum from
Cygnus X-1 corresponding to 78.9 minutes EOT between MJD
54002.928 and 54002.987 (2006-09-24). Also shown are the Crab nebula
spectrum and the best fit of a power-law to the data and the 95$\%$
CL upper limits to the steady $\gamma$-ray flux.}
\label{fig:spectrum}
\end{figure}

Cygnus X-1 was observed with MAGIC \cite{Goebel2007} between June and
November 2006 \cite{Albert2007b}. The data set comprises 40.0 good
observation hours from 26 different nights (see Table~\ref{table:log}
for details). A search for {\it steady} $\gamma$-ray signals was
performed for the entire recorded data sample, yielding no significant
excess. This allows us to establish the first upper limits to the VHE
$\gamma$-ray steady flux of Cygnus X-1 in the range between 150 GeV
and 3 TeV (see Figure~\ref{fig:spectrum}), of the order of $\leq
1--5\%$ of the Crab nebula flux. Given the timescale of the
variability of Cygnus X-1 in other energy bands, $\gamma$-ray signals
were also searched for on a daily basis. The results are shown in
Table~\ref{table:log}. We obtain results compatible with background
fluctuations at 99$\%$ CL for all searched samples except for
MJD=54002.875 (2006-09-24). We derive upper limits to the integral
flux above 150 GeV between 2 and 25$\%$ of the Crab nebula flux
(depending basically on the observation time) for all samples
compatible with background fluctuations. The data from 2006-09-24 was
further subdivided into two halves to search for rapidly varying
signals, obtaining 0.5$\sigma$ and 4.9$\sigma$ results for the first
(75.5 minutes EOT starting at MJD 54002.875) and second (78.9 minutes
EOT starting at MJD 54002.928) samples, respectively. The post-trial
probability is conservatively estimated by assuming 52 trials (2 per
observation night) and corresponds to a significance of 4.1$\sigma$.
The sample corresponding to MJD 54002.928 was further subdivided into
halves, obtaining 3.2$\sigma$ and 3.5$\sigma$ excesses in each. At
this point we stopped the data splitting process.

\begin{figure}[!t]
\begin{center}
\includegraphics [width=0.4\textwidth]{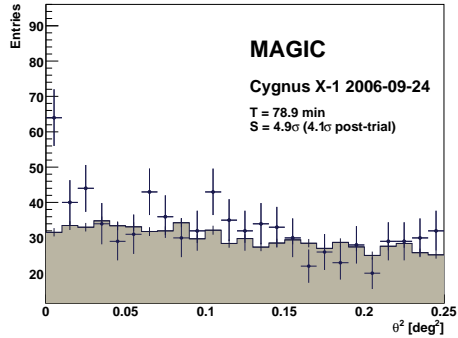}
\end{center}
\vspace{-0.5cm} 
\caption{Distribution of $\theta^2$ values for
the signal (dots) and background (histogram) events for an energy
threshold of 150 GeV.\label{fig:theta2}}
\end{figure}

\begin{figure}[!t]
\begin{center}
\includegraphics [width=0.4\textwidth]{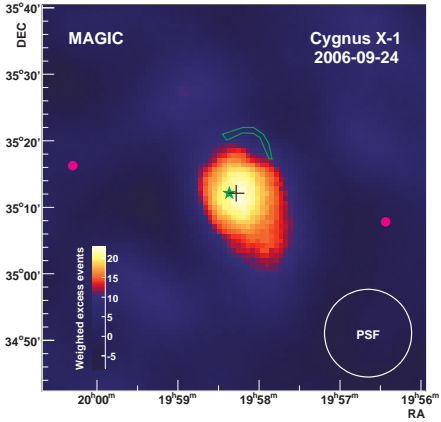}
\end{center}
\vspace{-0.5cm} 
\caption{Gaussian-smoothed ($\sigma = 4'$) map of
$\gamma$-ray excess events (background subtracted) above 150 GeV
around Cygnus X-1 corresponding to 78.9 minutes EOT between MJD
54002.928 and 54002.987 (2006-09-24). The black cross shows the
best-fit position of the $\gamma$-ray source. The position of the
X-ray source and radio emitting ring are marked by the green star
and contour, respectively. The purple dots mark the directions tracked
during the observations. Note that the bin contents are correlated due
to the smoothing.}
\label{fig:skymap}
\end{figure}

The distribution of $\theta2$ for signal and background events
corresponding to the 78.9 minutes EOT sample starting at MJD 54002.928
is shown in Figure~\ref{fig:theta2}. The excess is consistent with a
point source located at the position of Cygnus X-1. The map of excess
events around the source is shown in Figure~\ref{fig:skymap}. A
Gaussian fit yields the location: $\alpha = 19^\textrm{h}
58^\textrm{m} 17^\textrm{s}$, $\delta = 35^\circ 12' 8''$ with
statistical and systematic uncertainties of $1.5^\prime$ and $2'$,
respectively, compatible within errors with the position of Cygnus X-1
and excluding the radio nebula at a distance of $\sim 8'$. The energy
spectrum is shown in Figure~\ref{fig:spectrum}. It is well fitted
($\chi2/n.d.f=0.5$) by the following power law: $dN/(dA~dt~dE) =
(2.3\pm0.6)\times 10^{-12} (E/1~\textrm{TeV})^{-3.2\pm 0.6}
\textrm{cm}^{-2} \textrm{s}^{-1} \textrm{TeV}^{-1}$ where
the quoted errors are statistical only. We estimate the systematic
uncertainty to be $35\%$ on the overall flux normalization and 0.2 in
the determination of the spectral index.

\begin{figure}[!t]
\begin{center}
\includegraphics [width=0.45\textwidth]{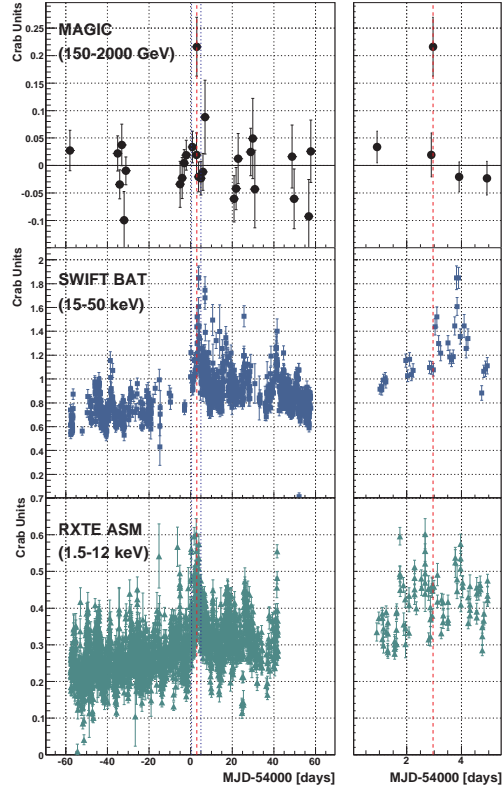}
\end{center}
\vspace{-0.5cm} 
\caption{From top to bottom: MAGIC, {\it Swift}/BAT
and {\it RXTE}/ASM measured fluxes from Cygnus X-1 as a function of
time. The left panels show the whole time spanned by MAGIC
observations. The vertical, dotted blue lines delimit the range zoomed
in the right panels. The vertical red line marks the time of the MAGIC
signal.}
\label{fig:lightcurve}
\end{figure}

The excess from the direction of Cygnus X-1 occurred simultaneously
with a hard X-ray flare detected by {\it INTEGRAL} \cite{Turler2006},
{\it Swift}/BAT and {\it RXTE}/ASM. Figure~\ref{fig:lightcurve} shows
the correlation between MAGIC, {\it Swift}/BAT and {\it RXTE}/ASM
light-curves. The TeV excess was observed on the rising side of the
first hard X-ray peak, 1--2 hours before its maximum, while there is
no clear change in soft X-rays. Additionally, the MAGIC {\it
non-detection} during the following night (yielding a $95\%$ CL upper
limit corresponding to a flux $\sim$5 times lower than the one
observed in the second half of 2006-09-24) occurred during the decay
of the second hard X-ray peak. A possible explanation is that, during
the night of 2006-09-24, soft and hard X-rays were produced in
different regions. Furthermore, hard X-rays and VHE $\gamma$-rays
could be produced in regions linked by the collimated jet, e.g.\ the
X-rays at the jet base and $\gamma$-rays at an interaction region
between the jet and the stellar wind. These processes would have
different physical timescales, thus producing a shift in time between
the TeV and X-ray peaks. Note that the distance from the compact
object to the TeV production region is constrained below 2$^\prime$ by
MAGIC observations and therefore it is unrelated with the nearby radio
emitting ring-like structure \cite{Gallo2005}. The observed TeV excess
took place at phase 0.91 (phase 0.0 is when the BH is behind the O
star). Currently, MAGIC observations are available only for the night
2006-09-24, which precludes any possible analysis of periodicity for
the TeV emission. The jet scenario, however, has some constraints. If
the TeV emission were produced in the jet, well within the binary
system, the photon-photon conversion in the stellar radiation field
would be very substantial, renderiing unlikely a TeV
detection\cite{Bednarek2007}. Admittedly, the inclination of the orbit
and the angle of propagation to the observer can affect this. Even
without an explanation for a TeV flare, it is possible that the
emission could have originated far from the compact
object. Interactions of the jet with the stellar wind may lead to such
a situation.

In summary, for the first time we have found experimental evidence of
VHE emission produced by a Galactic stellar-mass BH. It is also the
first evidence of VHE gamma-rays produced at an accreting binary
system. Our results show that a possible steady VHE flux is below the
present IACT's sensitivity and tight upper limits have been
derived. On the other hand, we find evidence for an intense flaring
episode during the inferior conjunction of the optical star, on a
timescale shorter than 1 day with a rise time of about 1 hour,
correlated with a hard X-ray flare observed by {\it Swift} and {\it
INTEGRAL}. These results imply the existence of a whole new
phenomenology in the young field of VHE astrophysics of binary systems
to be explored by present and future IACT's.

We thank the IAC for the excellent working conditions at the
Observatorio del Roque de los Muchachos in La Palma.

\bibliography{icrc0551}
\bibliographystyle{plain}
\end{document}